\documentstyle[12pt,epsfig]{article}
\newcommand{\pmb}[1]{%
        \setbox0=\hbox{#1}%
        \kern-.02em\copy0\kern-\wd0
        \kern+.04em\copy0\kern-\wd0
        \kern-.02em\raise.0217em\box0}
\newcommand{\vek}[1]{
        \mathchoice{\mbox{\boldmath$#1$}}%
        {\mbox{\boldmath$#1$}}%
        {\pmb{$\scriptstyle#1$}}%
        {\pmb{$\scriptscriptstyle#1$}}}

\newcommand{\lsim}{
 \mathrel{\setbox0=\hbox{$<$}\raise0.6ex\copy0\kern-\wd0
 \lower0.65ex\hbox{$\sim$}}}

\newcommand{\gsim}{
 \mathrel{\setbox0=\hbox{$>$}\raise0.6ex\copy0\kern-\wd0
 \lower0.65ex\hbox{$\sim$}}}

\begin{document}

\begin{titlepage}
\vspace*{-2cm}
\begin{flushright}
\bf 
TUM/T39-96-19
\end{flushright}

\bigskip 

\begin{center}
{\large \bf Diffractive $\rho^0$ photo- and leptoproduction 
            at high energies $^{*)}$}

\vspace{2.cm}

{\large G. Niesler, G. Piller  and W. Weise}  

\bigskip

Physik Department, Technische
Universit\"at M\"unchen
\\ D-85747 Garching, Germany\\ 

\vspace{5.cm}

{\bf Abstract}

\medskip
\begin{minipage}{15cm}
We discuss the elastic photo-  and leptoproduction  of 
$\rho^0$-mesons from nucleons at $Q^2\lsim 1 \,GeV^2$ 
as studied in recent experiments at HERA and FNAL. 
We find that the mass distribution of the measured $\pi^+\pi^-$ pairs 
is determined to a large extent by the two-pion 
contribution to the photon spectral function as given 
by the pion form factor. 
With rising $Q^2$ the rate of diffractive events decreases and  
the $\pi^+\pi^-$ mass distribution approaches a symmetric shape.

\end{minipage}

\end{center}

\vspace*{1.cm}


\vspace*{2.cm}

{\sl \centerline {To be published in Phys. Lett. B}}

\vspace*{2.cm}

\noindent $^{*}$) Work supported  in part by BMBF.

\end{titlepage}

The elastic photo- and leptoproduction of $\rho^0$-mesons from nucleons 
at high energies and moderate momentum transfers $Q^2 \lsim 1\,GeV^2$ 
has been actively discussed for many years. 
Earlier data are available from fixed target experiments at 
a  photon-nucleon center of mass energy $W \lsim 20 \,GeV$ 
(for a review see \cite{Baueea78}). 
There is a renewed interest in such processes because of  
recent collider measurements at HERA taken at 
$Q^2 \lsim 2\cdot10^{-2}\,GeV^2$  
and $W\simeq 180 \,GeV$  \cite{ZEUS94,ZEUS95}. 
Cross sections for $\rho^0$ leptoproduction at 
$Q^2 \lsim 1\,GeV^2$ are presently deduced from recent 
data taken by the E665 collaboration at Fermilab \cite{E665}.
In those experiments $\pi^+ \pi^-$ pairs are detected with an invariant mass 
$M_{\pi \pi} < 1\,GeV$ at small transverse momenta. 
They result dominantly from the  
$\rho^0 \rightarrow \pi^+ \pi^-$ decay \footnote{
In the following we identify always $\rho$ with the 
neutral $\rho^0$-meson and drop the index.};
a small fraction of the measured $\pi^+ \pi^-$ pairs 
is due to the production of uncorrelated pions.
Contributions from $\omega$- and $\phi$-mesons 
(apart from $\rho-\omega$ mixing) are excluded.

At high energies the elastic photoproduction of $\pi^+\pi^-$ pairs  
is characterized  by a weak energy dependence  and  
an exponential decrease of the production cross section 
with $t$, the squared four-momentum transfer. 
Furthermore, the vector meson is observed to retain, to a good 
approximation, the helicity of the incoming photon.
Such features are typical of  diffractive processes 
in high energy hadron-hadron collisions. 
This similarity can be understood by looking at the space-time 
structure of the high energy photon-nucleon interaction 
in the laboratory frame, where the target is at rest. 
Here the dominant contribution to the interaction cross section 
results  from  processes 
in which the photon fluctuates to a hadronic Fock state, 
in presence of the nucleon target,  and subsequently
scatters diffractively from this target \cite{Baueea78}.

Given the new and upcoming data, we find it useful to present 
in this note an improved and updated study of the mass distribution 
of diffractively produced  
$\pi^+ \pi^-$ pairs in the $\rho$-resonance region.
It was observed already long   ago \cite{Baueea78} that this   
mass distribution is skewed compared to a Breit-Wigner shape: 
there is an enhancement at the low mass side and a suppression 
of large mass contributions above the resonance. 
For an explanation several models were proposed 
(see  e.g. \cite{Soedin66,RosSto66}), mostly based on 
the interference of non-resonant and resonant $\pi^+ \pi^-$ 
contributions as a  source for the observed  asymmetry.

Since then a more detailed understanding of the 
$\rho$-meson and its coupling to the low mass $\pi^+ \pi^-$ continuum 
has been reached in terms of an effective field theory  which 
approximates QCD in the region of composite hadrons 
(see e.g. \cite{KlKaWe96} and references therein).
We point out that the mass distribution of $\pi^+ \pi^-$ pairs 
in diffractive photoproduction is  consistent, up to small corrections,
with the one deduced  from  the pion form factor in the  
timelike momentum region.
The latter is known  to  high precision from 
$e^+ e^- \rightarrow \pi^+ \pi^-$ annihilation.
In this process resonant and non-resonant $\pi^+ \pi^-$ states are 
automatically accounted for, and there is no need to separate 
them, as we shall demonstrate.

To describe the diffractive photoproduction 
of low mass pion pairs, $M_{\pi \pi} < 1\,GeV$,  
we apply a  generalized vector meson dominance model. 
Within this framework the corresponding cross section reads 
\cite{SakSch72}:
\begin{equation} \label{eq:D2cross_a}
\frac{d^2\sigma_{\gamma N\rightarrow \pi^+ \pi^- N}}{dM_{\pi\pi}^2 dt}
= 
\frac{\alpha}{4} \,\frac{\Pi\!\left(M_{\pi\pi}^2\right)}
                          {M_{\pi\pi}^2} \, 
\frac{\left| T_{\pi\pi N}(W^2,t) \right|^2}
{W^4},
\end{equation}
with $\alpha = 1/137$. Here $\Pi\!\left(M_{\pi\pi}^2\right)$ represents 
the two-pion contribution to the photon spectral function as measured in   
$e^+ e^- \rightarrow \pi^+ \pi^-$ annihilation: 
\begin{equation} \label{eq:PI}
\Pi\!\left(M_{\pi\pi}^2\right) = 
\frac{1}{12 \pi^2} \frac{\sigma_{e^+ e^- \rightarrow \pi^+ \pi^-}}
                        {\sigma_{e^+ e^- \rightarrow \mu^+ \mu^-}}.
\end{equation}
It accounts for the probability that the photon fluctuates 
into a $\pi^+\pi^-$ pair with invariant mass $M_{\pi\pi}$ 
in presence of the target. 
The subsequent scattering of the pion pair is 
described by the amplitude 
$T_{\pi \pi N}$. 
This amplitude is normalized such that its imaginary part is related 
to the effective $\pi \pi$-nucleon cross section as usual by the optical
theorem, $\sigma_{\pi\pi N} = Im T_{\pi\pi N}(W^2,t=0)/W^2$,
at large energies $W$.  
The picture is that of a high energy $\pi\pi$ pair acting like 
a ``beam'' as it scatters from the nucleon. 
We neglect inelastic contributions (e.g. those involving 
components with more than two pions in the 
photon spectral function) in this so-called diagonal approximation.

At the large center of mass energies of interest here 
one can safely neglect the  
real part of the $\pi^+ \pi^-$--nucleon forward scattering amplitude 
\cite{DonLan89} and finds:
\begin{equation} \label{eq:D2cross_b}
\left.\frac{d^2\sigma_{\gamma N\rightarrow \pi^+ \pi^- N}}{dM_{\pi\pi}^2 dt}
\right|_{t\approx 0}  = 
\frac{\alpha}{4} \,\frac{\Pi\!\left(M_{\pi\pi}^2\right)}
                          {M_{\pi\pi}^2} \,\sigma_{\pi\pi N}^2. 
\end{equation}
The effective $\pi^+\pi^-$--nucleon cross section $\sigma_{\pi\pi N}$ 
can in principle depend on the invariant $\pi \pi$ mass $M_{\pi \pi}$. 
We take $\sigma_{\pi\pi N}$ to be a constant, 
averaged over all (resonant and non-resonant) pion pairs 
with $M_{\pi\pi}< 1\,GeV$. 
Its magnitude is expected to be  of the order of the empirical  
$\rho$--nucleon total cross section, 
$\sigma_{\rho N} \sim (20-30)\,mb$ \cite{Baueea78}.

From experiment \cite{Baueea78,ZEUS94,ZEUS95} it is known that diffractive 
amplitudes decrease exponentially,  $d\sigma/dt \sim e^{b t}$, with 
the squared momentum transfer $t<0$.  
The diffractive excitation of a $\pi^+\pi^-$ pair with invariant 
mass $M_{\pi\pi}$ requires a non-zero longitudinal 
momentum transfer $k_L$.
In the laboratory frame with the $z$-axis chosen along the 
photon momentum,  $q = (\nu, \vek {0_\perp}, \nu)$, one finds  
$k_L = M_{\pi\pi}^2/2\nu $. 
At large photon energies $\nu \gsim 100\,GeV$ the squared 
minimal momentum transfer 
$t_{min} \approx  -k_L^2 = - M^4_{\pi\pi}/4\nu^2$
is negligible and 
we obtain for the $t$-integrated differential cross section:
\begin{equation} \label{eq:Dcross}
\frac{d\sigma_{\gamma N\rightarrow \pi^+ \pi^- N}}{dM_{\pi\pi}} = 
\frac{\alpha}{2 b } \,\frac{\Pi\!\left(M_{\pi\pi}^2\right)}
                          {M_{\pi\pi}} \,\sigma_{\pi\pi N}^2.
\end{equation}
Note  that the observed energy dependence of 
the diffractive photoproduction cross section,  
$d\sigma/d M_{\pi\pi} \sim W^{4(\alpha(0)-1)} \approx W^{0.32}$  
\cite{ZEUS95}, 
translates into an energy dependence of 
the effective $\pi\pi N$ cross section,  
$\sigma_{\pi\pi N} \sim W^{2 (\alpha(0) -1)} \approx 
W^{0.16}$. 
Such a behavior is similar to the energy dependence of 
total cross sections in high energy hadron-hadron collisions, 
usually described by the exchange of a ``soft'' pomeron 
with Regge-intercept $\alpha(0) \approx 1.08$ \cite{DonLan89}.

As an aside we 
mention that the diffractive cross section in 
eq.(\ref{eq:D2cross_b}) also describes successfully the shadowing contribution 
from low mass $\pi^+ \pi^-$ pairs in high-energy photon-deuteron scattering
\cite{PiRaWe95}. 
To verify this, note that the total photon-deuteron cross section  
has  two contributions: 
the single scattering contribution which gives 
twice the photon-nucleon cross section, 
and the double scattering term in which the photon interacts 
coherently with both proton and neutron in the deuteron.
This latter process reduces the total photon-deuteron 
cross section as compared to twice the photon-nucleon cross section,
i.e. it causes  shadowing. 
Its contribution to the total photon-deuteron cross section 
is related to the differential cross section 
$\frac{d^2\sigma_{\gamma N\rightarrow X N}}{dM_X^2 dt}$
for the diffractive photoproduction of hadrons with invariant 
mass $M_X$ from free nucleons (see e.g. \cite{Gribov69}): 
\begin{equation} \label{eq:shad}
\sigma^{(2)} = - 4\pi \int_{4 m_{\pi}^2}^{W^2} 
dM_X^2 
\left. \frac{d^2\sigma_{\gamma  N\rightarrow X N}}{dM_X^2 dt}
\right|_{t\approx 0} 
{\cal F}_d\left(k_L\right).
\end{equation}
The diffractive excitation of heavy hadronic states $X$   
requires a large momentum transfer $k_L$,   
which is however suppressed by the longitudinal deuteron form factor 
${\cal F}_d$.
Replacing the diffractive cross section in (\ref{eq:shad}) 
by (\ref{eq:D2cross_b}) yields 
the contribution of low mass $\pi^+ \pi^-$ pairs which dominates 
the shadowing effect \cite{PiRaWe95}.

To proceed further a good representation of the 
$\pi^+ \pi^-$ contribution to the photon spectral function 
(\ref{eq:PI}) is needed.
The latter is related to the pion form factor $F_{\pi}$ 
as follows:
\begin{equation} \label{eq:PiFpi}
\Pi\!\left(M_{\pi\pi}^2\right) = \frac{1}{48 \pi^2} 
\Theta\!\left( M_{\pi \pi}^2 - 4 m_{\pi}^2 \right) 
\left(1- \frac{4 m_{\pi}^2}{M_{\pi\pi}^2}\right)^{3/2} 
\left| F_{\pi}\left(M_{\pi\pi}^2\right)\right|^2.
\end{equation}
At 
timelike four-momenta the pion form factor is dominated by 
the $\rho$-meson resonance. 
An improved representation of the pion form factor, 
derived recently \cite{KlKaWe96},
gives perfect agreement with the measured form factor.  
It is based on an  
effective Lagrangian which combines vector meson dominance and chiral 
dynamics. The result is \cite{KlKaWe96}:
\begin{equation} \label{eq:piff1}
F_{\pi}(q^2) = \left( 1 - \frac{g_{\rho\pi\pi}}{g_{\rho}(q^2)} 
\frac{q^2}{q^2 - m_{\rho}^2 + i m_{\rho} \Gamma_{\rho}(q^2)}
\right)
\left( 1 + \frac{g_{\rho}(q^2)}{g_{\omega}}  
\frac{z_{\rho\omega}}{q^2 - m_{\omega}^2 + i m_{\omega} \Gamma_{\omega}}
\right). 
\end{equation}
The first term in (\ref{eq:piff1}) involves the dominant 
$\rho$-meson contribution.
The  width  
\begin{equation} 
\Gamma_{\rho} = \frac{g^2_{\rho \pi\pi}}{48 \pi m_{\rho} \sqrt{q^2}} 
(q^2 - 4 m_{\pi}^2)^{3/2} 
\end{equation}  
reflects the strong coupling of the $\rho$-meson to 
the $\pi^+\pi^-$ continuum, with $g_{\rho\pi\pi} = 6.05$.   
The effective $\gamma\rho$ coupling with inclusion of  vertex 
corrections due to the $\pi\pi$ loop is: 
\begin{equation} 
\frac{1}{g_{\rho}(q^2)} \equiv \frac{1}{\stackrel{o}{g}_{\rho}} - 
\frac{m_{\rho}^2 - \stackrel{o}{m}_{\rho}^2 - i \,
m_{\rho}\,\Gamma_{\rho}(q^2)}
{g_{\rho \pi \pi} \,q^2}. 
\end{equation} 
The bare and physical $\rho$-meson masses are   
$\stackrel{o}{m}_{\rho} = 0.81\,GeV$ and  
$m_{\rho} = 0.775\,GeV$, respectively. 
Their difference comes from the real part of the $\rho\rightarrow \pi\pi$ 
self-energy as explained in details in ref.\cite{KlKaWe96}. 
The bare $\gamma\rho$ coupling 
constant is  fixed as $\stackrel{o}{g}_{\rho} = 5.44$ 
to reproduce the $\rho\rightarrow e^+ e^-$ partial width.
The second term in eq.(\ref{eq:piff1}) yields  a fine tuning of 
$F_{\pi}$ due to $\rho-\omega$-mixing, with  
$m_{\omega} = 0.782\,GeV$, $g_{\omega} = 17.0$, 
$\Gamma_{\omega} = 8.4\,MeV$ and 
$z_{\rho\omega} = -4.52 \cdot 10^{-3} \,GeV^2$.
In Fig.1 the resulting pion form factor from (\ref{eq:piff1}) 
is shown together with data from ref.\cite{Barkea85}. 
The agreement is evidently very satisfactory.

With eqs.(\ref{eq:Dcross},\ref{eq:PiFpi},\ref{eq:piff1}) 
we have 
calculated the $\pi^+\pi^-$ mass distribution $d\sigma/dM_{\pi\pi}$.
For a comparison with recent data of the ZEUS collaboration 
\cite{ZEUS95}, taken at an average  center of mass energy
$\overline W = 70 \, GeV$,  
we use the measured  value of  the slope parameter $b \approx 10 \,GeV^{-2}$.
With an  effective $\pi^+\pi^-$--nucleon cross section  
$\sigma_{\pi\pi N} = 30\, mb$ the main features of 
the observed mass distribution are reproduced as shown in Fig.2. 
We conclude that the mass distribution of  
diffractively produced  $\pi^+\pi^-$ pairs 
is indeed determined primarily by the two-pion component of 
the photon spectral function, or equivalently, by the pion form 
factor in the timelike momentum region.
Note again that the latter already includes a substantial contribution 
from non-resonant pion pairs.
Corrections due to a possible  mass dependence of $\sigma_{\pi\pi N}$,  
or from  non-diagonal  inelastic corrections, are evidently small. 
To quantify  their possible size we fit the experimental 
data by adding the following correction to the diffractive cross section 
(\ref{eq:Dcross}):
\begin{equation} \label{eq:fit}
\sigma_{\pi\pi N}^2 \rightarrow \sigma_{\pi\pi N}^2\left( 1 + 
c\, \frac{m_{\rho}^2 - M_{\pi\pi}^2}{M_{\pi\pi}^2} \right),
\end{equation}
with c = 0.6. 
This fit is shown as the dashed curve in Fig.2. 
The correction in (\ref{eq:fit}) is substantially smaller 
than in previous fits to $\pi^+ \pi^-$ photoproduction data 
(see \cite{ZEUS95,DonMir87} and references therein). 
The reason for this improvement is mainly due to a proper 
treatment of the energy dependent width of the $\rho$-meson 
as discussed in detail in \cite{KlKaWe96}.

Finally we extend our considerations to  diffractive  
leptoproduction processes at moderate $Q^2=-q^2 \lsim 1\,GeV^2$. 
It is an empirical fact that in this kinematic region 
the exclusive leptoproduction  of $\rho$-mesons from nucleons 
is well described within the  vector meson dominance 
picture \cite{Baueea78}, which gives: 
\begin{equation} \label{eq:DcrossQ2}
\frac{d\sigma_{\gamma^* N \rightarrow \pi^+\pi^- N}}{dM_{\pi\pi}}(Q^2)
= 
\frac{d\sigma_{\gamma N \rightarrow \pi^+\pi^- N}}{dM_{\pi\pi}}(Q^2=0)
\left(\frac{M_{\pi\pi}^2}{M_{\pi\pi}^2 + Q^2}\right)^2 
\left(1 + \epsilon \,\xi^2 \frac{Q^2}{M_{\pi\pi}^2}\right).
\end{equation}
Here $\xi$ is the ratio of the longitudinal to transverse 
$\pi^+\pi^-$--nucleon forward amplitudes and $\epsilon$ measures the 
longitudinal polarization of the virtual photon.  
We investigate  the diffractive cross section   
(\ref{eq:DcrossQ2}) in the kinematic range  of 
recent measurements performed by the E665 collaboration at 
Fermilab \cite{E665}.
Here the average center of mass energy is $\overline W = 15 \,GeV$. 
Consequently we have to re-scale the previously determined value for 
the effective $\pi^+\pi^-$--nucleon cross section: 
$(\sigma_{\pi\pi N})_{E665} \approx  
(\sigma_{\pi\pi N})_{ZEUS}\cdot(W_{E665}/W_{ZEUS})^{0.16} \approx 24\,mb$.  
Furthermore we use $\xi^2 = 0.5$ \cite{Baueea78} and $\epsilon = 0.9$. 
\cite{E665}. 
In Fig.3 we show the results for the diffractive cross 
section (\ref{eq:DcrossQ2}) for different values of $Q^2$,  
including  the correction term from (\ref{eq:fit}). 
As expected, the diffractive cross section decreases rapidly 
with rising $Q^2$. Furthermore the mass distribution 
approaches a more symmetric shape as $Q^2$ increases.

Finally we study  the ratio of the diffractive to the inelastic 
scattering cross section at moderate $Q^2$: 
\begin{equation} \label{eq:diff/dis}
\frac{\sigma_{diff}}{\sigma_{inel}} = 
\frac{1}{\sigma_{\gamma^* N}} \,
{\int_{4 m_{\pi}^2}^{1\,GeV^2} dM_{\pi\pi}
\frac{d\sigma_{\gamma^*N\rightarrow \pi^+\pi^- N}} {dM_{\pi\pi}}}.
\end{equation}
At $W=15\,GeV$ and $Q^2 = 0$ we obtain  
${\sigma_{diff}}/{\sigma_{inel}} \approx 0.1$. 
With increasing $Q^2$ we observe a decrease of diffractive 
events compared to inelastic ones. 
For example at $Q^2 = 1 \, GeV^2$ we find 
${\sigma_{diff}}/{\sigma_{inel}} \approx 0.06$.

In summary, we find that the mass distribution of $\pi^+\pi^-$ 
pairs in diffractive photoproduction is determined to a large 
extent by the two-pion contribution to the photon spectral 
function as given  by the pion form factor. 
Corrections due to a possible mass dependence of the effective 
$\pi^+\pi^-$--nucleon 
cross section, or from non-diagonal inelastic processes, are small.
When applied to diffractive leptoproduction at moderate $Q^2$ 
we observe  a rapid decrease of the production cross section, while 
the mass distribution of $\pi^+\pi^-$ pairs approaches a symmetric 
shape with rising $Q^2$. 
As $Q^2$ increases we find a  decrease of 
diffractive events as compared to inelastic ones.

\bigskip
\bigskip
\noindent
We would like to thank W. Wittek and T.J. Carroll for helpful 
discussions and comments.

\newpage

{\bf Figure Captions}

\begin{itemize}

\item[Figure 1:]
The pion form factor $F_{\pi}(q^2)$ in the region of timelike 
$q^2$. The data are from ref.\protect{\cite{Barkea85}}. The solid line 
shows the result using eq.(\protect{\ref{eq:piff1}}).

\item[Figure 2:]
The mass distribution $d\sigma/dM_{\pi\pi}$. 
The data are from ref.\protect{\cite{ZEUS95}}.  
The solid line is the result using eq.(\protect{\ref{eq:Dcross}}). 
The dashed line includes the correction in eq.(\protect{\ref{eq:fit}}).

\item[Figure 3:]
The mass distribution $d\sigma/dM_{\pi\pi}$ for different 
values of $Q^2$ for $W=15\,GeV$ and $\epsilon = 0.9$. 

\end{itemize}

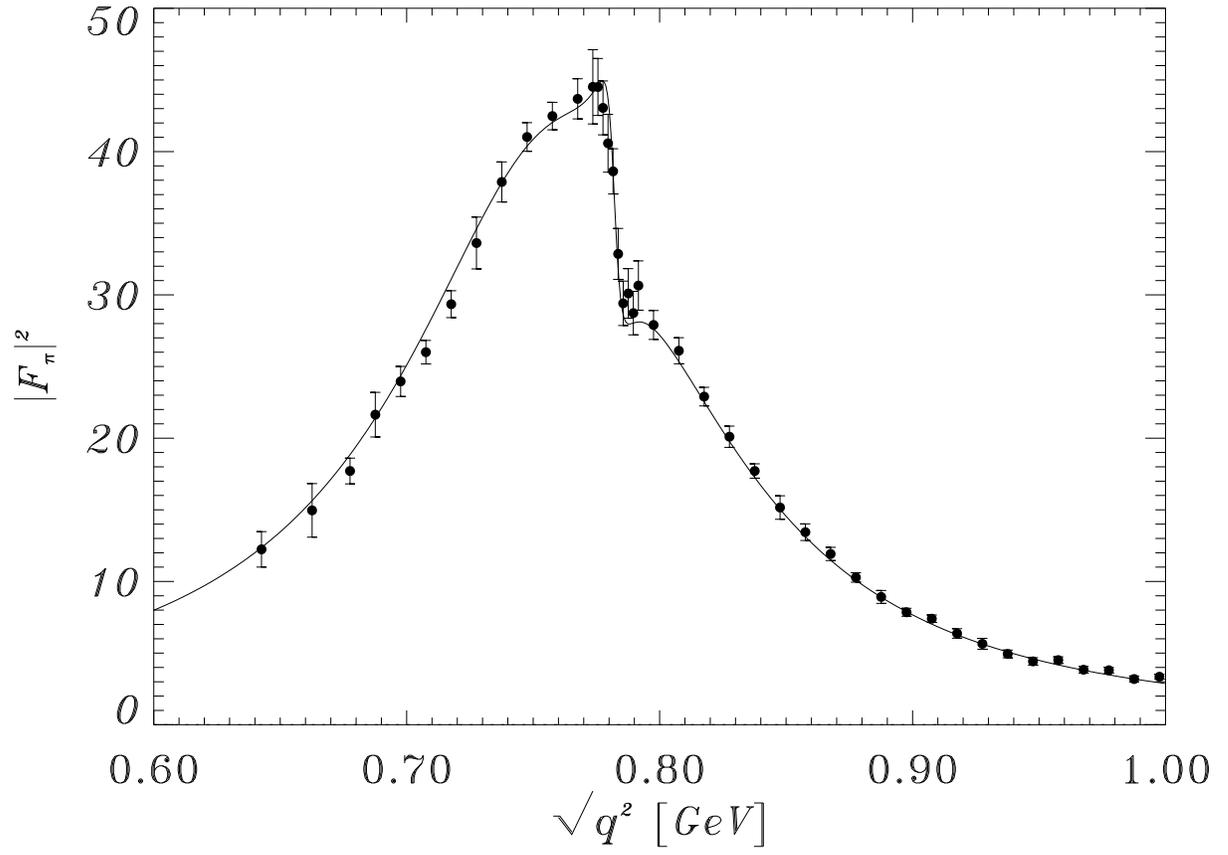
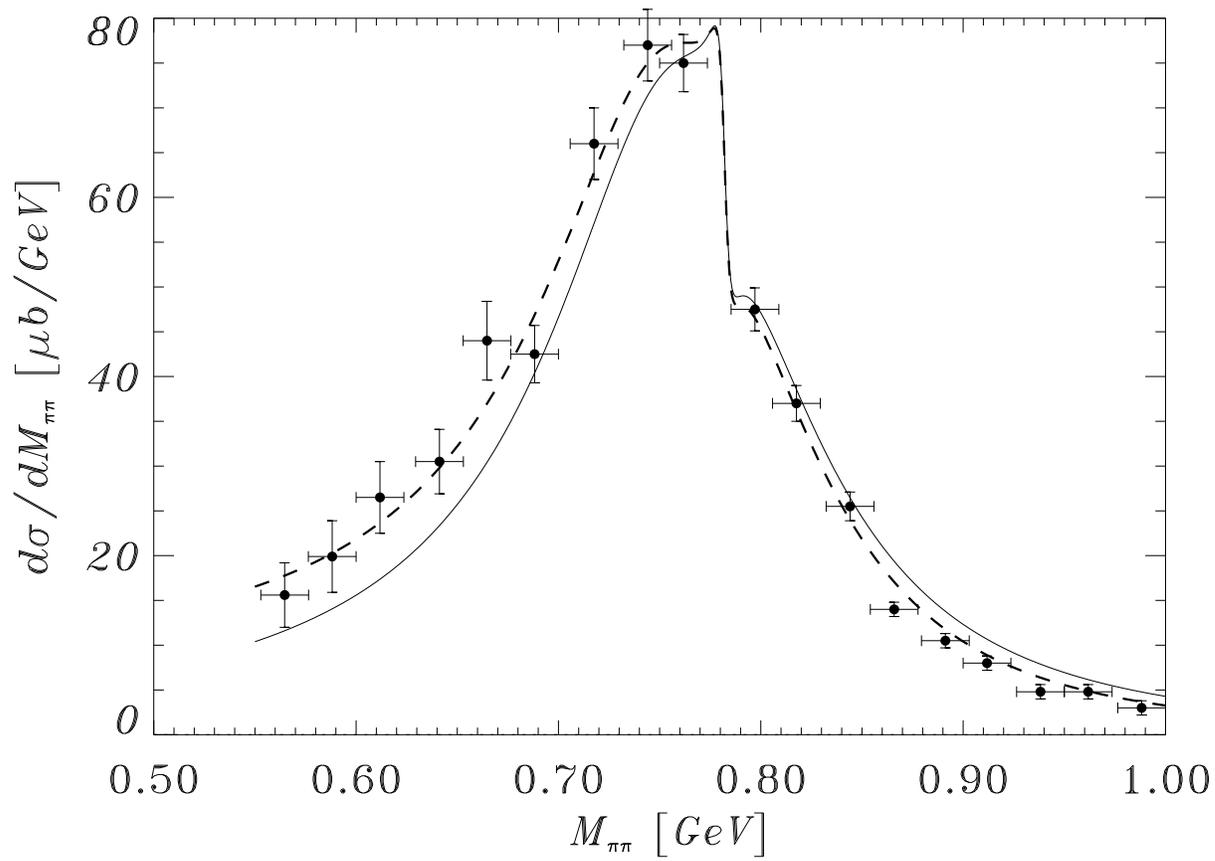
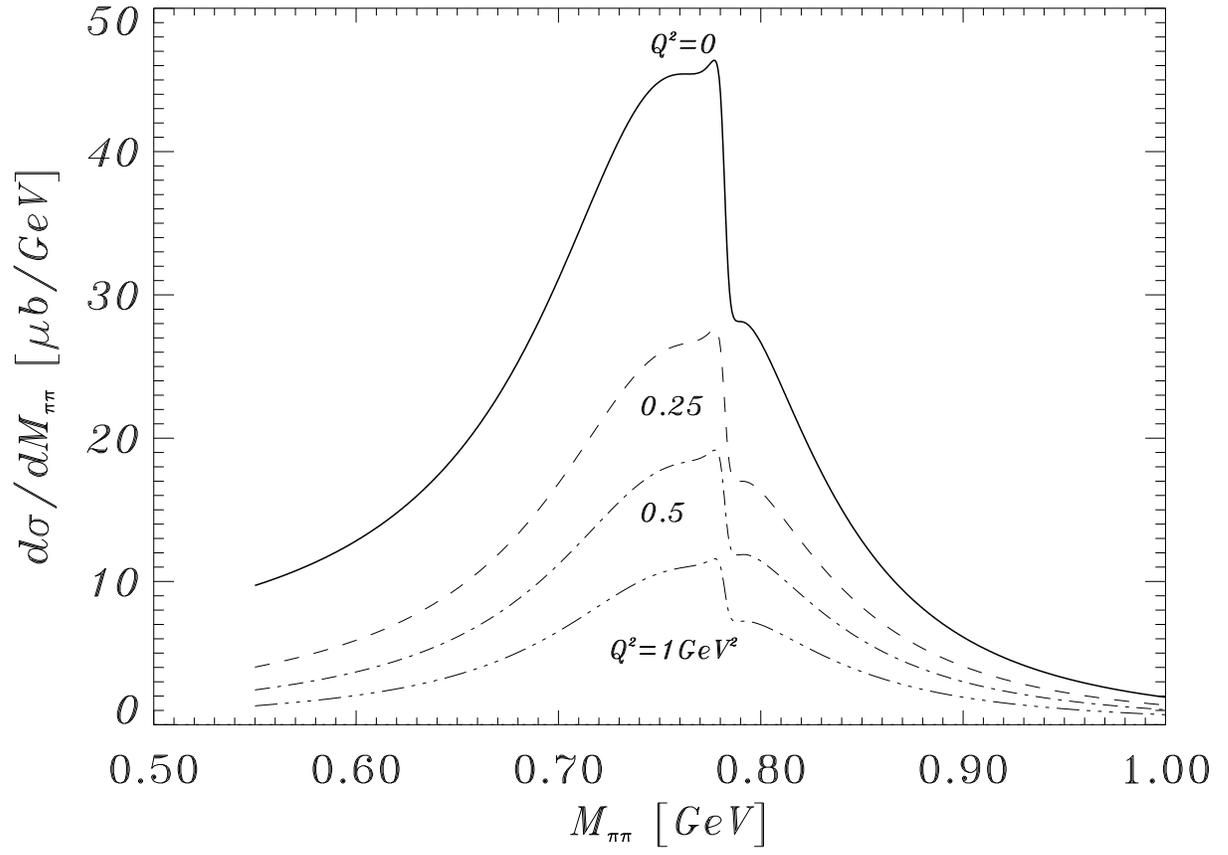
\end{document}